# Frozen discord in stochastic dephasing environment

**Bin Yi**(易斌)[1]

[1]*University College London*

Abstract: Preserving quantum coherence is fundamental challenge in the field of quantum computation. Here, I investigate the frozen discord phenomenon and non-Markovianity for qubits experiencing local dephasing in a classical environment. The quantum discord of a certain initial bipartite state, independently interacting with stochastic Gaussian fields, can be frozen for a finite time. The preservation of discord is shown to be intimately related to the backflow of information from the environment. The relevant result for quantum dephasing environment is compared.

## 1. Introduction

Quantum correlations among components of composite systems are essential for quantum information processing [33, 34]. Realistic systems are open and unavoidably interact with the surrounding environment, destroying the quantum features of the system after a certain amount of time [23-32]. This result especially manifests in a configuration of independent qubits locally interacting with their environment [35], which is the standard setting for quantum communication tasks [34]. Detrimental effects from the environment are generally divided into two types. The first type is decoherence due to entanglement between the system and its environment [37]. The system and bath are considered to be a closed system with an evolution governed by an overall unitary operation. The second type involves the introduction of classical noise into the system when we do not have full control over the applied fields. Both types of noise are common in nature, but they have distinct origins.

Recently, there has been some interest in designing viable procedures to protect quantum correlations against detrimental noise from the environment [11-14, 23-32]. In particular, quantum discord, which is a quantum correlation measure that is more general than entanglement, has been extensively studied recently [9,10]. Quantum discord could be exploited to provide an advantage over classical computations [36]. In the past decade, the phenomenon of frozen discord was discovered [11-14]. For a certain class of initial bipartite states locally interacting with dephasing environments, the discord between the two qubits could be preserved for a certain amount of time. Quantum dephasing environments were considered in Ref. [14], and the authors discovered that the discord of the system could even remain at a constant value forever at experimentally reachable low temperatures. This preservation and revival [23-32] of the quantum correlation of a quantum system has been closely associated with non-Markovianity, which measures the backflow of information from the environment to the system [16-22].

In many situations, a full quantum description of the environment is unfeasible, while classical stochastic modeling provides a valid alternative. In this article, I investigate the frozen discord phenomenon when a bipartite system is locally subject to Gaussian stochastic noise. The paper is structured as follows: In Section 2, the classical model is reviewed. Section 3 discusses how frozen discord depends upon the noise parameters of the classical dephasing channel. Section 4 studies the relationship between frozen discord and non-Markovianity, and Section 5 compares the classical and quantum dephasing environments.

## 2. Classical dephasing model

Consider a harmonic oscillator with natural frequency $\omega_0$ subject to stochastic classical fields with time-dependent fluctuating complex amplitude $B(t)$ and central frequency $\omega$. Assuming qubits are represented by the number states of a harmonic oscillator, the Hamiltonian, in units of $\hbar$, is given by

1) $H = \omega_0 a^\dagger a + (a\overline{B}(t)e^{i\omega t} + a^\dagger B(t)e^{-i\omega t})$

where the first term is the free Hamiltonian term, and the term in the parentheses is the interaction Hamiltonian.

In the interaction picture, the Hamiltonian reads:

2) $H_I(t) = ae^{-i\delta t}\overline{B}(t) + a^\dagger e^{i\delta t}B(t)$

where $\delta = 1 - \omega$ (in units of $\omega_0$) is the detuning between the natural frequency of the system and the central frequency of the classical stochastic fields.

Within this rotating frame, the time evolution operator is given by:

3) $U(t) = \tau \exp[-i\int_0^t ds H_I(s)]$

where $\tau$ denotes time ordering and can be dropped since $[H_I(t_1), H_I(t_2)] \propto I$, with $I$ being the identity operator.

We may then write the evolution of the system density matrix as follows:

4) $\rho(t) = [U(t)\rho(0)U^\dagger(t)]_B = [D(\phi_t)\rho(0)D^\dagger(\phi_t)]_B$

where $D(\mu) = e^{\mu a^\dagger - \bar\mu a}$ is the displacement operator, $\phi_t = -i\int_0^t ds e^{i\delta s} B(s)$ is the time-dependent displacement of the argument, and $[...]_B$ denotes an average over a stochastic process.

Consider a stochastic field $B(t) = B_x(t) + iB_y(t)$ with real and imaginary parts that independently obeys a Gaussian distribution with a mean of zero [2].

5) $[B_x(t)]_B = [B_y(t)]_B = 0$

6) $[B_x(t_1)B_x(t_2)]_B = [B_y(t_1)B_y(t_2)]_B = K(t_1, t_2)$

7) $[B_x(t_1)B_y(t_2)]_B = [B_y(t_1)B_x(t_2)]_B = 0$

where $K(t_1, t_2)$ is the kernel autocorrelation function.

The stochastic average in Eq. (4) can then be computed. The evolution of the system, expanded in terms of Fock states, is given by a Gaussian channel [1]:

8) $\rho(t) = \sum_{m,n} \rho_{nm} e^{-\frac{1}{2}(n-m)^2 \beta(t)} |n\rangle\langle m|$

where $|n\rangle$ represents a Fock state, $\rho_{nm}$ is the matrix element of the initial state of the system and $\beta(t)$, which plays the role of the variance of the Gaussian channel, is defined as

9) $\beta(t) = \int_0^t \int_0^t ds_1 ds_2 \cos[(s_1 - s_2)\delta] K(s_1, s_2)$

The evolution of a single qubit is represented by Eq. (8) with $n$ and $m$ taking the value of 0 or 1. The diagonal terms of the density matrix remain unchanged, while the off-diagonal terms are suppressed by a factor of $e^{-\frac{1}{2}\beta(t)}$.

For the Ornstein-Uhlenbeck process, an important example of a stationary Gaussian process, the kernel autocorrelation function is given by:

10) $K_{OU}(t_1, t_2) = \frac{\lambda}{2t_E} \exp(-\frac{|t_1 - t_2|}{t_E})$

where $\lambda$ is the coupling constant, and $t_E$ is the characteristic time of the environment. Substituting Eq. (10) into Eq. (9) gives:

11) $\beta_{OU}(t) = \frac{\lambda}{(1+(\delta t_E)^2)^2}\{t - t_E + (\delta t_E)^2(t + t_E) + t_E e^{-t/t_E}[(1-(\delta t_E)^2)\cos \delta t - 2\delta t_E \sin \delta t]\}$

As $t_E$ approaches 0,

12) $\beta_{OU}(t) = \lambda t + \lambda t_E e^{-t/t_E} \cos \delta t$

Therefore, the stochastic model reduces to a phasing diffusion channel in the case where the correlation time goes to zero, with the coupling constant $\lambda$ equal to the dephasing rate.

Eq. (11) can be rescaled in units of $t_E$ by letting $\tilde{\delta} = \delta t_E$, $\tilde{\lambda} = \lambda t_E$ and $\tilde{t} = t/t_E$, leading to

13) $\beta_{OU}(t) = \dfrac{\tilde{\lambda}}{(1+(\tilde{\delta})^2)^2} \{\tilde{t} - 1 + \tilde{\delta}^2(\tilde{t}+1) + e^{-\tilde{t}}[(1-\tilde{\delta}^2)\cos\tilde{\delta}\tilde{t} - 2\tilde{\delta}\sin\tilde{\delta}\tilde{t}]\}$

For simplicity, this rescaled expression of $\beta(t)$ will be used throughout the following chapters. Note that a large value of $\tilde{\delta}$ produces oscillations if it exceeds a threshold value [3]

14) $\tilde{\delta}_0 = \dfrac{3\pi}{2}[productLog(\dfrac{3\pi}{2})]^{-1} = 3.644$

The oscillatory behavior of the variance $\beta(t)$ in the stochastic dephasing model is closely related to quantum revival [3-8]. As the value of $\tilde{\delta}$ increases, the number of oscillations of $\beta(t)$ increases, while the magnitude of $\beta(t)$ becomes small. In the dephasing model, $\beta(t)$ is closely related to the noise added to the system. A high value of the detuning $\tilde{\delta}$ prevents the damaging action of the environment. The role of the detuning parameter in correlation preservation and the backflow of information will be thoroughly discussed in the following chapters.

## 3. Frozen discord of a two-level system in classical environments

Quantum discord is a measure of quantum correlations. It goes beyond the distinction between entangled and separable states [9]. For a bipartite state, the quantum discord between the two states is defined as $Q(\rho_{AB}) = I(\rho_{AB}) - C(\rho_{AB})$, where $I(\rho_{AB})$ is the mutual information given by $I(\rho_{AB}) = S(\rho_A) + S(\rho_B) - S(\rho_{AB})$, with $S(\rho) = -Tr(\rho \log(\rho))$ being the von Neumann entropy. $C(\rho_{AB}) = \max_{\Pi_A} J(\Pi_A \rho_{AB})$ represents classical correlations [9,10], with $\Pi_A$ being the orthogonal projection operator acting on qubit A, and $J$ quantifies the information gained about the system as a result of the measurement $\Pi_A$.

Recently, it was discovered that the discord of a certain class of initial bipartite states interacting independently with local dephasing environments can be frozen for a finite amount of time [11-14] while its classical correlation decays. This frozen discord phenomenon occurs for Bell-like initial states of the form

15) $\rho_{AB} = \dfrac{1+c}{2}|\Psi^\pm\rangle\langle\Psi^\pm| + \dfrac{1-c}{2}|\Phi^\pm\rangle\langle\Phi^\pm|$

where $|\Psi^\pm\rangle$ and $|\Phi^\pm\rangle$ are the four Bell states, and $|c|<1$.

In this article, I study how the detuning parameter in the classical model affects the sudden transition between the classical and quantum decoherence regimes. Moreover, I study whether discord can also be frozen forever if the qubits are locally subject to Gaussian stochastic fields, with the model given by Section II. For the initial state given by Eq. (15), the mutual information and classical correlations take the following form:

16) $I(\rho_{AB}(t)) = \sum_{j=1}^{2} \dfrac{1+(-1)^j c}{2}\log_2[1+(-1)^j c] + \sum_{j=1}^{2} \dfrac{1+(-1)^j c}{2} e^{-\Lambda(t)} \log_2[1+(-1)^j e^{-\Lambda(t)}]$

17) $C[\rho_{AB}(t)] = \sum_{j=1}^{2} \dfrac{1+(-1)^j \chi(t)}{2}\log_2[1+(-1)^j \chi(t)]$

where $\chi(t) = \max\{e^{-\Lambda(t)}, c\}$, $\Lambda(t)$ is the dephasing factor and c is taken to be positive for simplicity. The specific form of $\Lambda(t)$ in the stochastic field model is $\dfrac{1}{2}\beta(t)$, and the rescaled expression of $\beta(t)$ for the Ornstein-Uhlenbeck process is given by Eq. (13)

Observing Eq. (16) and Eq. (17), one finds that for $e^{-\Lambda(\bar{t})} > c$, quantum discord $Q(\rho_{AB}) = I(\rho_{AB}) - C(\rho_{AB})$ is given by the constant term in Eq. (16), while classical correlations decay. For $e^{-\Lambda(\bar{t})} < c$, quantum discord decays and classical correlations remain constant. Therefore, if a finite time t satisfies

18) $e^{-\Lambda(\bar{t})} = c$,

then $\bar{t}$ is the transition time that splits the classical decoherence and quantum decoherence regions.

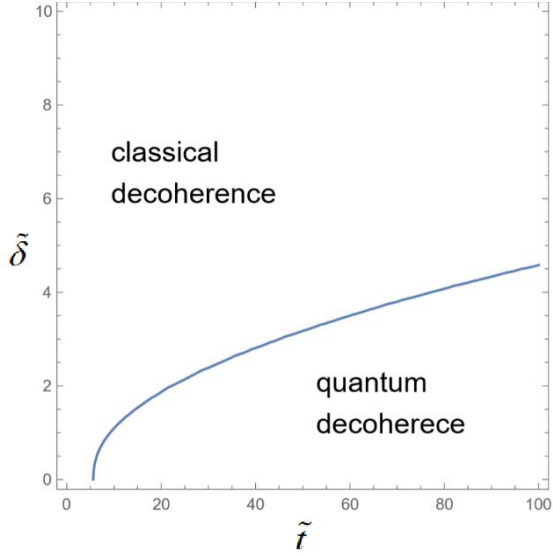

Figure 1: Contour plot of the correlation dynamics in the $\tilde{\delta}-\tilde{t}$ plane for $c=0.1$ and $\tilde{\lambda}=1$. In the classical decoherence region (left), classical correlations decay and discord remains constant. In the quantum decoherence region (right), quantum discord decays and classical correlations remain constant. The transition between the two regions corresponds to the values of $\tilde{\delta}$ and $\tilde{t}$ such that Eq. (18) is satisfied.

Figure 1 shows the values of $\tilde{\delta}$ and $\tilde{t}$ such that condition Eq. (18) is satisfied. The transition time increases as $\tilde{\delta}$ increases. In other words, the frozen discord time is prolonged. For instance, the discord of the initial state, measured in entropic units (*e.u.*) with $c=0.1$, is given by $0.007225$ (*e.u.*). The discord stays at this value until it reaches the transition time and then starts decaying. The transition time for the resonant scenario, where $\tilde{\delta}=0$, is given by $\tilde{t} \cong 5.6$, as shown Figure 2. For $\tilde{\delta}=10$, $\tilde{t} \cong 463.2$.

The rescaled detuning parameter $\tilde{\delta}$ has also been demonstrated to shield the evolution of the system from detrimental actions of the environment in other information processing tasks [1,3,15]. This result is expected, since $\tilde{\delta}$ is the product of the detuning parameter and the correlation time, both of which are inner-correlated with non-Markovianity. An open system is dissipative because information and energy can flow to the environment. However, information can also flow from the environment back to the system, leading to non-Markovian fluctuations. $t_E$ quantifies the amount of time the environment can store the memory. Therefore, a large value of $t_E$ means that quantum correlations in the system can be better preserved. The relation between frozen discord and non-Markovianity will be further discussed in the next chapter.

Finally, notice that in the limit $\lim_{t\to\infty} \beta(t) = \infty$, the variance $\beta(t)$ shown in Eq. (13) diverges. Therefore, a transition time always exists given a value of $c$. In other words, the discord of the bipartite state locally interacting with the classical environment cannot be preserved indefinitely.

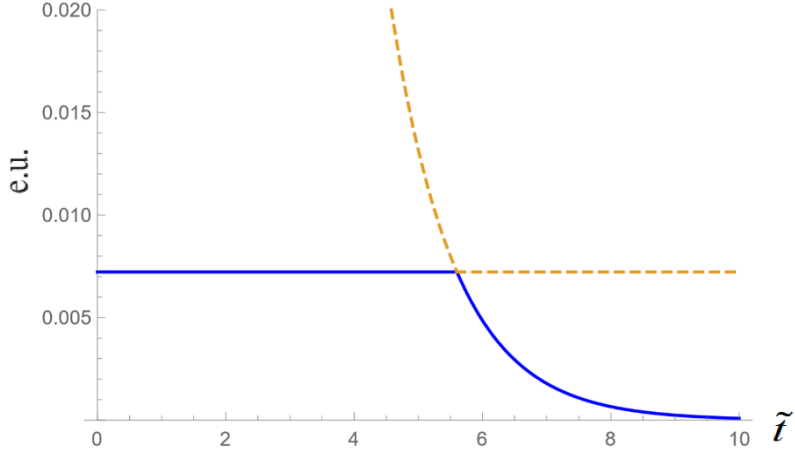

Figure 2: Dynamics of discord (solid blue) and classical correlations (dashed orange), measured in entropic units ( $e.u.$ ) as a function of $\tilde{t}$ for $c = 0.1$, $\tilde{\delta} = 0$ and $\tilde{\lambda} = 1$.

## 4. Frozen discord and non-Markovianity

If a quantum system is monitored by an environment continuously, the quantum information encoded in the system will be lost. However, in certain situations, the lost information can flow back to the system due to correlations between the system and the environment. The system is said to be non-Markovian if such recoherence occurs. Various measures of non-Markovianity have been proposed recently, attempting to quantify the amount of information flowing back to the system [16-22]. In particular, I will focus on the quantum capacity measure [16]. This measure links non-Markovian dynamics with an increase in the efficiency of quantum information processing. This measure of non-Markovianity based on the nonmonotonic behavior of quantum capacity reads

$$19)\ N_Q = \int_{\frac{dQ(\Phi_t)}{dt}>0} \frac{dQ(\Phi_t)}{dt} dt$$

where $Q(\Phi_t)$ is the quantum capacity of channel $\Phi_t$, which gives the maximum rate at which quantum information can be reliably sent down a quantum channel and is defined in terms of coherent information between the input and output of the quantum channel $\Phi_t$. The specific form of $Q$ for the dephasing channel acting on a single qubit is given by

$$20)\ Q^D(t) = 1 - H_2\left(\frac{1+e^{-2\Lambda(t)}}{2}\right)$$

where $H_2$ is the binary Shannon entropy. For N qubits interacting with independent and identical environments, the dephasing channel is degradable, regardless of the specific structure of the environment. Hence, the capacity measure is additive. For the setting considered here, two qubits in independent environments, no new features are present compared to the single qubit scenario, up to a renormalization factor. Therefore, Eq. (20) will be used in the following discussions for simplicity.

The quantum capacity Eq. (20) for different values of $\tilde{\delta}$ is plotted in Figure 3. Note that no oscillation occurs for $\tilde{\delta} < \tilde{\delta}_0 = 3.644$ [3]; therefore, $N_Q(\tilde{\delta} < 3.644) = 0$. The non-Markovianity measure $N_Q$ for different values of $\tilde{\delta}$ is shown in Figure 4. For instance, $N_Q(\tilde{\delta} = 5) = 0.021059$ ( e.u. ) and $N_Q(\tilde{\delta} = 10) = 0.047582$ ( e.u. ). For $\tilde{\delta} > 3.644$, the non-Markovianity increases as $\tilde{\delta}$ increases.

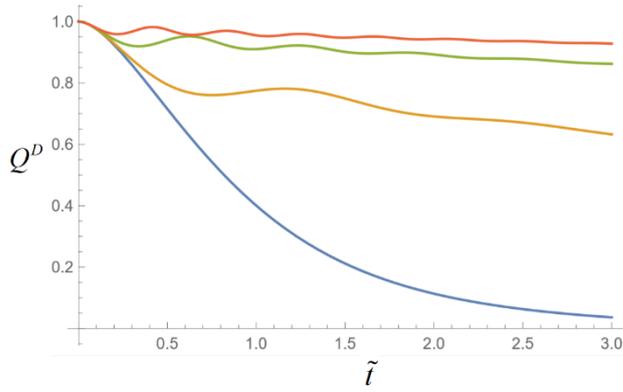

Figure 3: Quantum capacity of the stochastic dephasing model as a function of $\tilde{t}$ for different values of $\tilde{\delta}$ with $\tilde{\lambda} = 1$. The curves from bottom to top corresponds to $\tilde{\delta} = 1$ (blue), $\tilde{\delta} = 5$ (orange), $\tilde{\delta} = 10$ (green), and $\tilde{\delta} = 15$ (red)

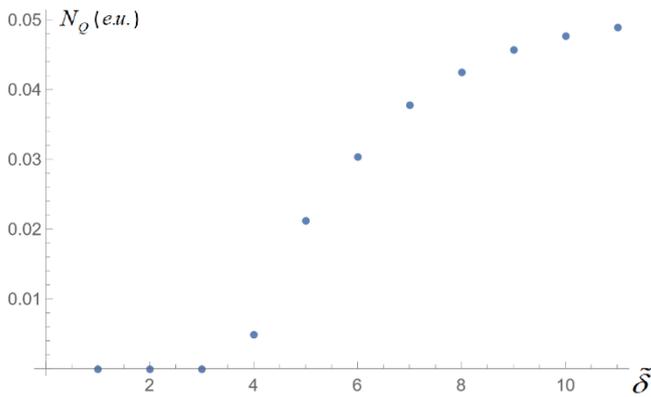

Figure 4: Non-Markovianity measure $N_Q$ in entropic units for values of $\tilde{\delta}$ ranging from 1 to 11.

Comparing Figure 1 and Figure 4, one finds that the frozen discord phenomenon occurs for both Markovian and non-Markovian dynamics. For $\tilde{\delta} = 3$, discord freezes for a time $\tilde{t} \approx 42$ while the non-Markovianity measure reads zero. The dynamics become nondissipative as $\tilde{\delta}$ approaches 0, where the frozen discord phenomenon still manifests. For values of $\tilde{\delta}$ larger than $\tilde{\delta}_0 = 3.644$, the freezing time is prolonged as the non-Markovianity increases.

## 5. Comparison with dephasing quantum environments

The interaction with quantum environments is, in general, more complicated. Each qubit can exchange quantum correlation with the reservoir in which it is embedded. For the same initial states considered in Eq. (15) locally subject to a quantum environment, the discord can remain at a constant value indefinitely. As shown in Ref. [14], each qubit interacts independently with a quantum reservoir with an Ohmic-like spectrum. The discord measure of the states $Q(\rho_{AB})$ can remain at a constant value not only for a finite time but also indefinitely for certain values of c and the Ohmicity parameter. Time-invariant discord and non-Markovianity have a similar dependence on the Ohmicity parameter. Both require the suppression of coupling to low-frequency modes. Therefore, regardless of whether the environment is classical or quantum, the frozen discord phenomenon is inner correlated with non-Markovianity, which measures the information backflow. Non-Markovianity indicates the capacity of an open system to recover its quantumness.

In dissipative quantum environments, the emergence of invariant or reviving quantum correlations has been explained in terms of quantum correlation exchange, including entanglement, among qubits and a quantum environment [23-28]. Memory effects lead to back-action of the quantum environments on the qubits. When qubits are subject to classical environments, such back-action is absent. The revival or invariance of quantum correlations in this case has been explained in terms of the backflow of classical information, and several interpretations have been proposed [29-32]. For instance, one could perceive the classical environment as a controller for which a unitary operation is acting on the system. The controller keeps a record of what action is applied to the system, and when the record is lost, the quantum correlations of the system are damaged; when the information is recovered, the quantum correlations revive [29, 31].

## 6. Conclusion and perspectives

In this article, I investigate the frozen discord phenomenon for bipartite states locally interacting with classical environments. The quantum discord of a certain class of initial states, locally subject to stochastic Gaussian fields, can remain at a constant value for a finite time. Moreover, the relationship between frozen discord and non-Markovianity, which measures the backflow of information from the environment to the system, is studied. Frozen discord is prolonged as the detuning parameter, and therefore the non-Markovianity, increases. This observation indicates that the preservation of quantum discord is inner correlated with the backflow of information from the environments to the system.

This article may inspire the implementation of invulnerable quantum communication devices. Since Gaussian classical noise is common in experiments [38-40], the results imply that if we exclude the input state to a certain set, the computational resource (quantum discord) is particularly resistant to detrimental effects from the environment until a transition time.

Acknowledgment

The author would like to thank Anthony J. Leggett (adviser) for fruitful discussions and constant encouragements.